\documentclass[prl,twocolumn,
longbibliography,
showpacs,
aps
]{revtex4-2}
\usepackage{graphicx}
\usepackage{amsmath}
\usepackage{amsfonts}
\usepackage{amssymb}
\usepackage[colorlinks=true,linkcolor=blue,urlcolor=blue,citecolor=blue]{hyperref}
\usepackage{color}
\usepackage{lipsum} 
\newcommand{\ket}[1]{|{#1}\rangle}
\newcommand{\bra}[1]{\langle{#1}|}

\newcommand{\beq}{\begin{equation}}
	\newcommand{\eeq}{\end{equation}}

\newcommand{\sx}{\hat{b}_{\uparrow}^{\dag}\hat{b}_{\downarrow}^{\phantom{\dag}}+\hat{b}_{\downarrow}^{\dag}\hat{b}_{\uparrow}^{\phantom{\dag}}}

\usepackage{dsfont}
\usepackage{placeins} 
\usepackage{bm}           

\usepackage{scalerel}
\usepackage{tikz}
\usetikzlibrary{svg.path}
\definecolor{orcidlogocol}{HTML}{A6CE39}
\tikzset{
	orcidlogo/.pic={
		\fill[orcidlogocol] svg{M256,128c0,70.7-57.3,128-128,128C57.3,256,0,198.7,0,128C0,57.3,57.3,0,128,0C198.7,0,256,57.3,256,128z};
		\fill[white] svg{M86.3,186.2H70.9V79.1h15.4v48.4V186.2z}
		svg{M108.9,79.1h41.6c39.6,0,57,28.3,57,53.6c0,27.5-21.5,53.6-56.8,53.6h-41.8V79.1z M124.3,172.4h24.5c34.9,0,42.9-26.5,42.9-39.7c0-21.5-13.7-39.7-43.7-39.7h-23.7V172.4z}
		svg{M88.7,56.8c0,5.5-4.5,10.1-10.1,10.1c-5.6,0-10.1-4.6-10.1-10.1c0-5.6,4.5-10.1,10.1-10.1C84.2,46.7,88.7,51.3,88.7,56.8z};
	}
}
\newcommand\orcid[1]{\href{https://orcid.org/#1}{\mbox{\scalerel*{
				\begin{tikzpicture}[yscale=-1,transform shape]
					\pic{orcidlogo};
				\end{tikzpicture}
			}{R}}}}

\begin{document}
	\title{Dissipative Dicke time crystals: an atoms' point of view}
	 \author{Simon~B.~J\"ager\,\orcid{0000-0002-2585-5246}}
	 \author{Jan Mathis Giesen}
	 \author{Imke Schneider\,\orcid{0000-0002-3820-0089}}
 	 \author{Sebastian Eggert\,\orcid{0000-0001-5864-0447}}
 	  \affiliation{Physics Department and Research Center OPTIMAS, University of Kaiserslautern-Landau, D-67663, Kaiserslautern, Germany}
	\begin{abstract}
	We develop and study an atom-only description of the Dicke model with time-periodic couplings between atoms and a dissipative cavity mode. The cavity mode is eliminated giving rise to effective atom-atom interactions and dissipation. We use this effective description to analyze the dynamics of the atoms that undergo a transition to a dynamical superradiant phase with macroscopic coherences in the atomic medium and the light field. Using Floquet theory in combination with the atom-only description we provide a precise determination of the phase boundaries and of the dynamical response of the atoms. From this we can predict the existence of dissipative time crystals that show a subharmonic response with respect to the driving frequency. We show that the atom-only theory can describe the relaxation into such a dissipative time crystal and that the damping rate can be understood in terms of a cooling mechanism. 
\end{abstract}
\maketitle
Time-periodic driving of quantum systems allows for the creation of tailored out-of-equilibrium structures including quantum states with topological order~\cite{Dalibard:2011,Wintersperger:2020} and self-organized coherent patterns~\cite{Staliunas:2002,Chin:2010,Perego:2016,Clark:2017,Nguyen:2019,Zhang:2020, Fazzini:2021}. Here, one distinguishes between off-resonant driving and resonant driving. In the high frequency limit, the former results in a quasi static quantum system that experiences an engineer-able and time-averaged Hamiltonian~\cite{Eckardt:2017,Weitenberg:2021}. Resonant driving, instead, enables strong dynamical coherences between otherwise weakly coupled quantum states which can force the quantum system in exotic spatio-temporal patterns. This is exciting as it allows the controlled, on-demand generation of purpose-oriented quantum states but comes at the cost ofdealing with an energetically open system, which requires a full understanding of relaxation and decoherence mechanisms to avoid heating by 
using engineered dissipation~\cite{Petiziol:2022,Zhu:2019}.
Such 
active {\it open system control} is in fact one of the main challenges for technological progress in the design of quantum matter. While the Lindblad formalism works well for photonic systems, the description of dissipation in condensed matter with massive particles is often fitted by phenomenological models, with limited understanding and tunability.

We now want to analyze a strongly correlated model of atoms where dissipation is derived microscopically from the interactions with the environment in order to pave the way for
quantum state engineering far from equilibrium.  For the dissipative Dicke model significant progress has been made to eliminate the cavity in order to derive an effective atom-only master equation~\cite{Damanet:2019,Bezvershenko:2021} which is of Lindblad form~\cite{Jaeger:2022}, i.e.~the relaxation is directly linked to the interactions of the atoms with photons.  
However, it is so far unclear if this derivation is valid for time-dependent or periodically driven systems, which are
far from equilibrium where the dissipation of large amounts of energy is required.  We will derive the atom-only 
description for the time-periodic dissipative Dicke model which is of large fundamental and prototypical interest. Here, resonant periodic driving can induce the formation of subharmonic spatio-temporal patterns, a so called dissipative time crystal (DTC).  This phase is accompanied by superradiant light emission into the cavity and was recently the focus of several experimental and theoretical works~\cite{Chitra:2015,Iemini:2018,Gong:2018,Zhu:2019,Kessler:2021,Kongkhambut:2021,Kongkhambut:2022,Kosior:2023,Nie:2023}.

Photon elimination is highly non-trivial in this case~\cite{Curtis:2019,Palacino:2021,Kelly:2022} since a strong coupling to 
the cavity is crucial for two separate mechanisms~\cite{Ritsch:2013,Mivehvar:2021,Defenu:2021}:
(i) It mediates tuneable time-periodic atom-atom interactions which are essential for the pattern formation via parametric driving~\cite{Chitra:2015,Molignini:2018,Cosme:2019,Cosme:2023}.  
(ii) The cavity generates dissipation that is required for stabilizing the emerging patterns. In the limit of strong atom-photon interaction the usual approach is therefore to treat the dynamics of atoms and cavity on equal footing.  

As we demonstrate in this Letter the elimination of photons is nonetheless possible and highly successful in the prediction of the full time evolution and the non-equilibrium phase diagram, underlining the advantages of an atom-only description. Analytic predictions of the lower stability threshold and a full analysis of the
spectral features and gaps are now possible, hence paving the way for future engineering of tailored dynamic 
atomic models, which can be used as quantum simulators of complicated interacting systems.

\textit{Model}-- We consider the time-periodic dissipative Dicke model and eliminate the cavity in order to derive an effective atom-only Master equation which is of  Lindblad form~\cite{Jaeger:2022}.
The dynamics of the density operator $\hat{\rho}$ describing the atoms and one coupled cavity mode with loss rate $\kappa$ is governed by the master equation ($\hbar=1$)
\begin{equation}
\label{eq:Master}
\frac{\partial\hat{\rho}}{\partial t}=-i\left[\hat{H},\hat{\rho}\right]-\kappa(\hat{a}^{\dag}\hat{a}\hat{\rho}+\hat{\rho}\hat{a}^{\dag}\hat{a}-2\hat{a}\hat{\rho}\hat{a}^{\dag}).
\end{equation}
The coupling to $N$ two-level atoms that are driven by an external laser
is described by the Hamiltonian~\cite{Dimer:2007,Baumann:2010,Larson:2021}
\begin{equation}
\label{eq:H}\hat{H}=\delta_c\hat{a}^{\dag}\hat{a}+\Delta \hat{n}_\uparrow +\frac{g(t)}{\sqrt{N}}(\hat{a}+\hat{a}^{\dag})\left(\sx \right) ,
\end{equation}	
where $\delta_c$ is the detuning between the cavity resonance and the external laser drive, $\hat a^\dag$ and $\hat a$ are the cavity field creation and annihilation operators, and the product of bosonic operators $\hat{b}_{\uparrow}^{\dag}\hat{b}_{\downarrow}^{\phantom{\dag}}$ change one atomic state from the ground state $\ket{\downarrow}$ to 
the metastable excited state $\ket{\uparrow}$ of energy $\Delta$. The operators $\hat{n}_\uparrow=\hat{b}^{\dag}_{\uparrow}\hat{b}^{\phantom{\dag}}_{\uparrow}$ and $\hat{n}_\downarrow=\hat{b}^{\dag}_{\downarrow}\hat{b}^{\phantom{\dag}}_{\downarrow}$ measure the number of atoms in each state such that $N=\hat{n}_{\uparrow}+\hat{n}_{\downarrow}$. A modulation of the external driving laser leads to a time-periodic collective coupling $g(t)=g_0+g_1 \cos \omega t $, corresponding to two side-bands of the drive.  

\begin{figure}
	\center
	\includegraphics[width=1\linewidth]{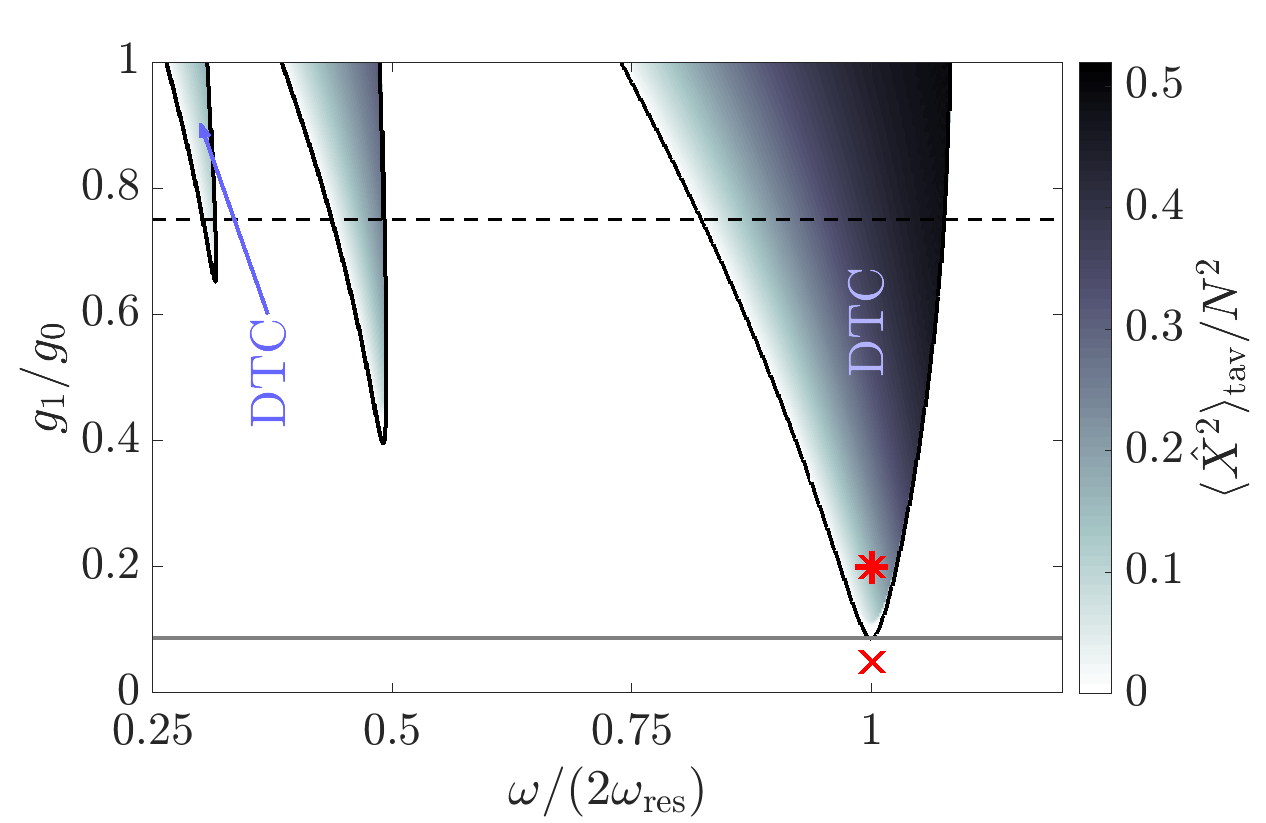}
	\caption{Time-averaged superradiance order parameter $\langle\hat{X}^2\rangle_{\mathrm{tav}}$ calculated from the mean-field equations~\eqref{eq:varphidownarrow},\eqref{eq:varphiuparrow} and evaluated at $\kappa t=10^4$ as function of the driving frequency $\omega/(2\omega_{\mathrm{res}})$ and modulation strength $g_1/g_0$. Solid lines mark the threshold to superradiance found by $\gamma_{\mathrm{max}}=0$. The horizontal gray solid line indicates the threshold $g_1^c/g_0$ given by Eq.~\eqref{g1c}. The red cross ($g_1=0.05g_0$, $\omega=2\omega_{\mathrm{res}}$) and star ($g_1=0.2g_0$, $\omega=2\omega_{\mathrm{res}}$) correspond to the parameters used in Fig.~\ref{Fig:2}(a) and (b), respectively. The dashed horizontal line shows the parameters visible in Fig.~\ref{Fig:3}, $g_1=0.75 g_0$. We used $\delta_c=\kappa$, $\Delta=0.1\kappa$, $g_0=0.5g_c$. DTC indicates where superradiant phases with subharmonic responses are found. \label{Fig:1}}
\end{figure}

In the static limit, $g_1=0$, the dissipative Dicke model [Eq.~\eqref{eq:Master}] shows a transition from  normal state to superradiance at $g=g_c=[\Delta(\delta_c^2+\kappa^2)/(4\delta_c)]^{1/2}$~\cite{DallaTorre:2013,Kirton:2019}. Superradiance is signaled by macroscopic coherences in the atomic medium $\langle\hat{X}^2\rangle\propto N^2$, $\hat{X}=\hat{b}_\uparrow^\dag\hat{b}_\downarrow+\hat{b}_\downarrow^\dag\hat{b}_\uparrow$, and a large cavity field $\langle\hat{a}^\dag\hat{a}\rangle\propto N$. In this Letter, we focus on the subcritical regime to study the influence of time-periodic driving with $g(t)<g_c$ at all times. In this situation, a dynamical superradiant configuration can still be found depending on the modulation strength $g_1$ when the driving frequency $\omega$ is close to a parametric resonance~\cite{Chitra:2015},  $n\omega=2\omega_{\mathrm{res}}$ ($n=1,2,\dots$) with resonance frequency \begin{align}
\omega_\mathrm{res}=\Delta\sqrt{1-\frac{g_0^2}{g_c^2}}.\label{omegares}
\end{align}
In Fig.~\ref{Fig:1} we show our results for the regions of superradiance in parameter space spanned by $g_1$ and $\omega$. The colorbar shows the time-averaged mean value of the superradiance order parameter $\langle\hat{X}^2\rangle_{\mathrm{tav}}=\int_{t}^{t+T}d\tau\langle\hat{X}^2(\tau)\rangle/T$. A derivation of this phase diagram is shifted to a later point in this Letter. This superradiant phase features a time-oscillatory superradiant order parameter $\langle\hat{X}^2\rangle$. DTC order appears if additionally the two-time correlation function $C_1(t,t_0)=\langle \hat{X}(t+t_0)\hat{X}(t_0)\rangle$ is periodic in $t$ with period $2T$, $T=2\pi/\omega$. This requires the breaking of a discrete time translational symmetry which happens whenever $n$ is odd (see DTC in Fig.~\ref{Fig:1}). The breaking of this symmetry implies the existence of a many-body mode oscillating with $\omega/2$ whose lifetime approaches infinity for increasing atom numbers. 

\textit{Atom-only description}-- First, we will derive the atom-only description, by extending the theory of Ref.~\cite{Jaeger:2022} and applying it for the first time to a time-dependent problem.
In the limit of a short cavity relaxation time, $|\delta_c-i\kappa|\gg\Delta,\omega,g$, we apply a
Schrieffer-Wolff transformation $\hat{D}(t)=\exp(\hat{a}^\dag\hat{\beta}(t)-\hat{\beta}^{\dag}(t)\hat{a})$ to eliminate the photonic degrees of freedom.  The condition for decoupling the atoms from the cavity modes in the master equation leads to a {\it time-dependent} equation of the transformation operators $\hat \beta(t)$
\begin{align}
i\frac{\partial\hat{\beta}}{\partial t}=(\delta_c-i\kappa)\hat\beta+\frac{g(t)}{\sqrt{N}}\left(\sx \right)+[\Delta \hat{n}_\uparrow,\hat{\beta}]\label{eq:beta}
\end{align}
which is solved by 
$\hat\beta(t) = c_+(t) \hat{b}_{\uparrow}^{\dag}\hat{b}_{\downarrow}^{\phantom{\dag}} + c_-(t) \hat{b}_{\downarrow}^{\dag}\hat{b}_{\uparrow}^{\phantom{\dag}}$ in the steady state. It is one major ingredient of this theory that we also include the commutator with $\Delta\hat{n}_\uparrow$ that adds retardation effects due to the time-evolution of the atoms. Without this term the dynamical stabilization of the atomic state is not possible. The resulting differential equation is discussed in the Supplemental Material (SM)~\cite{SM}, which yields an expansion
\begin{equation}
c_\pm (t)\approx -\frac{1}{\sqrt{N}}\left(\frac{g(t)}{\delta_c-i\kappa}+\frac{i\dot{g}(t)}{(\delta_c-i\kappa)^2}\mp
\frac{\Delta\,\,{g}(t)}{(\delta_c-i\kappa)^2}\right),
\label{coeff}
\end{equation}
where the first term corresponds to the quasi-static solution. 
With $\hat{\beta}$ we can then write the effective master equation for the atomic density operator $\hat{\rho}_{\mathrm{at}}=\mathrm{Tr}_{\mathrm{cav}}[\hat{D}^\dag\hat{\rho}\hat{D}]$ by tracing over the cavity degrees of freedom
\begin{equation}\label{eq:effmaster}
\frac{\partial \hat{\rho}_{\mathrm{at}}}{\partial t}=-i\left[\hat{H}_{\mathrm{at}},\hat{\rho}_{\mathrm{at}}\right]-\kappa(\hat{\beta}^{\dag}\hat{\beta}\hat{\rho}_{\mathrm{at}}+\hat{\rho}_{\mathrm{at}}\hat{\beta}^{\dag}\hat{\beta}-2\hat{\beta}\hat{\rho}_{\mathrm{at}}\hat{\beta}^{\dag}). 
\end{equation}
This atom-only description includes the coherent time evolution of the atoms governed by the Hamiltonian 
\begin{align}
\hat{H}_{\mathrm{at}}=&\Delta\hat{n}_\uparrow+\frac{g(t)}{2\sqrt{N}}\left(\hat{\beta}^\dag[\sx]+\mathrm{H.c.}\right). \label{bos}
\end{align}
The non-trivial time-dependence of $\hat\beta(t)$ therefore enters both the (i) cavity-mediated interactions in the second term of Eq.~\eqref{bos} and the (ii) cavity-generated dissipation $\propto \kappa$ in Eq.~\eqref{eq:effmaster}. In the SM~\cite{SM} we provide a comparison of the atom-cavity and atom-only theory described by Eqs.~\eqref{eq:Master} and \eqref{eq:effmaster}, respectively.

\begin{figure}
	\center
	\includegraphics[width=1\linewidth]{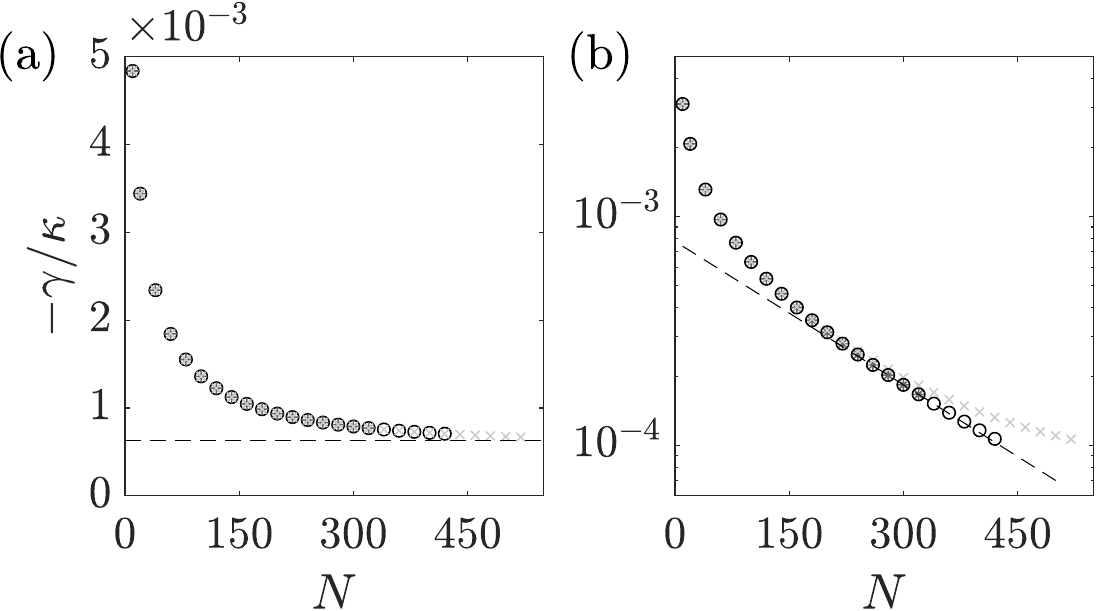}
	\caption{\label{Fig:2} Gap $\gamma$ in units of $\kappa$ as function of $N$ for (a) $g_1/g_0=0.05$ and (b) $g_1/g_0=0.2$. Different markers indicate different cut-offs in Floquet space (see SM~\cite{SM}, black circles $M_\mathrm{cut}=3$, gray stars $M_\mathrm{cut}=4$, light gray crosses $M_\mathrm{cut}=2$). The dashed line correspond to (a) the mean-field $\gamma_{\mathrm{Fl}}$ and (b) an exponential fit $\propto\exp(-0.005N)$.}
\end{figure}

\textit{Formation of a stable DTC}--The resulting atomic theory described by Eqs.~\eqref{eq:effmaster} and \eqref{bos} is a full quantum mechanical description of the dynamics of the atomic state, which is the main tool in this Letter. The massively reduced Liouville space dimension allows us to study spectral features of the time-crystalline phase for atom numbers that cannot be accessed with a full atom-cavity description. Given a time-periodic Liouvillian [\textit{i.e.} the right-hand side of Eq.~\eqref{eq:effmaster}] we can calculate the eigenmodes $\hat{\rho}_{\lambda}=e^{\lambda t}\hat{\varrho}_{\lambda}$ with a time-periodic $\hat{\varrho}_{\lambda}(t+T)=\hat{\varrho}_\lambda(t)$. The eigenvalues $\lambda$ have in general negative or zero real part, $\mathrm{Re}(\lambda)\leq0$, and because of the time-periodicity their imaginary value $\mathrm{Im}(\lambda)$ can be chosen within an interval of length $\omega$. The emergence of DTC is marked by the breaking of a discrete time-translational symmetry. This results in a closing gap $\gamma:=\mathrm{Re}(\lambda)$ in the spectrum for increasing atom number $N$ at a subharmonic frequency response $\mathrm{Im}(\lambda)=\omega/2$. The theory described by Eq.~\eqref{eq:effmaster} enables to our knowledge for the first time the study of a decreasing $|\gamma|$ in the time-periodic Dicke model with only atomic degrees of freedom. In Fig.~\ref{Fig:2}(a) and (b) we show $\gamma$ as function of $N$ (a) outside of the DTC phase (see red cross in Fig.~\ref{Fig:1}) and (b) in the DTC phase (see red star in Fig.~\ref{Fig:1}). In (a) we find that $\gamma$ converges to a constant highlighting that this mode remains gapped. In (b), instead, we find an exponential closing of the gap therefore indicating time-crystalline behavior (further details in SM~\cite{SM}). Our finding of an exponentially closing gap is consistent with previous claims~\cite{Gong:2018} and enabled by our atom-only description that can access atom numbers in a range which are elusive for a full quantum atom-cavity description. 

For very large atom numbers $N$ we can further simplify the full quantum description of the atom-only master equation to a mean-field description of $\varphi_s=\langle\hat{b}_s\rangle$, $s=\downarrow,\uparrow$ (see SM~\cite{SM})
\begin{align}
\frac{d\varphi_\downarrow}{dt}=&i\frac{V_0-iV_1}{N}|\varphi_\uparrow|^2\varphi_\downarrow+i\frac{V_0+iV_1}{N}\varphi_\uparrow^2\varphi_\downarrow^*,\label{eq:varphidownarrow}\\
\frac{d\varphi_\uparrow}{dt}=&-i\left(\Delta-\frac{V_0+iV_1}{N}|\varphi_\downarrow|^2\right)\varphi_\uparrow+i\frac{V_0-iV_1}{N}\varphi_\downarrow^2\varphi_\uparrow^*.\label{eq:varphiuparrow}
\end{align}
This mean-field description includes (i) coherent interactions and (ii) dissipation that are described as non-linear terms proportional to 
$V_0=-\sqrt{N}g(t)\mathrm{Re}(c_++c_-)$ and $V_1=N\kappa(|c_-|^2-|c_+|^2)$, respectively. The amplitude $c_-$ ($c_+$)  describe the likelihood of atoms undergoing a transition from $\ket{\uparrow}$ to $\ket{\downarrow}$ ($\ket{\downarrow}$ to $\ket{\uparrow}$). Note that $V_1\neq0$ is a consequence of including retardation effects described by the  commutator with $\Delta\hat{n}_\uparrow$ in Eq.~\eqref{eq:beta}. An imbalance, in our case $N(|c_-|^2-|c_+|^2)=4\delta_c\Delta  g^2(t)/(\delta_c^2+\kappa^2)^2>0$ for $\Delta,\delta_c>0$, leads to a preferential reduction of atomic excitations.  Consequently, dissipation described by $V_1$ has a nice physical interpretation: it is a cooling rate which is crucial for the stabilization of the system over long timescales. The efficient description given by Eqs.~\eqref{eq:varphidownarrow} and \eqref{eq:varphiuparrow} allows us to map out the whole phase diagram visible in Fig.~\ref{Fig:1}.

\textit{Threshold}--We will now show that it is possible to derive analytical results for the onset of superradiance.
We assume that all atoms are initially in the ground state and explore when driving induces an instability towards superradiance.  With most bosons in $\ket{\downarrow}$, we eliminate fluctuations in the ground state using $\varphi_\downarrow \approx \sqrt{N}$, which linearizes Eq.~\eqref{eq:varphiuparrow}. The resulting complex differential equation for $\boldsymbol{\varphi}_\uparrow=(\varphi_\uparrow,\varphi_\uparrow^*)$ can be solved using Floquet theory by making the ansatz $\boldsymbol{\varphi}_\uparrow(t)=e^{\lambda_\mathrm{Fl} t}{\bf u}(t)$ with a $T=2\pi/\omega$ periodic vector ${\bf u}$ and the Floquet eigenvalue $\lambda_\mathrm{Fl}=\gamma_\mathrm{Fl}-i\nu_\mathrm{Fl}$, $\gamma_\mathrm{Fl},\nu_\mathrm{Fl}\in\mathbb{R}$. Details of this derivation  are reported in the SM~\cite{SM}. The stability of the fluctuations is determined by $\gamma_{\mathrm{max}}$ which is the maximum of all possible real parts $\gamma_\mathrm{Fl}$. Whenever $\gamma_{\mathrm{max}}\leq0$ ($\gamma_{\mathrm{max}}>0$) we expect the system to be non-superradiant (superradiant). In the non-superradiant regime, the Floquet eigenvalues $\lambda_{\mathrm{Fl}}$ represent the low frequency modes $\lambda$ that are found using Floquet theory for the full Lindbladian in Eq.~\eqref{eq:effmaster} for $N\to\infty$. To demonstrate this we show $\gamma_{\mathrm{Fl}}$ as dashed line in Fig.~\ref{Fig:2}(a) which appears to be the thermodynamic limit of $\gamma$. Above threshold, for $\gamma_{\mathrm{Fl}}>0$, such a comparison is not possible as $\lambda_{\mathrm{FL}}$ can only describe the short-time dynamics.
The threshold to superradiance is marked by $\gamma_{\mathrm{Fl}}=0$ and shown as black line in Fig.~\ref{Fig:1}. To get analytical expressions, we reformulate the coupled complex differential equation as real second-order differential equation for $x_\uparrow=\varphi_\uparrow+\varphi_\uparrow^*$ 
\begin{align}
\frac{d^2x_\uparrow}{dt^2}+2V_1(t)\frac{dx_\uparrow}{dt}+\Delta\left[\Delta-2V_0(t)\right]x_\uparrow=0.\label{eq:secondorderdiff}
\end{align}
In this differential equation $V_0$ modifies the resonance frequency $\Delta$ originating from (i) cavity-mediated interactions and $V_1$ serves as a damping of fluctuations coming from (ii) the cavity-generated dissipation. 
If we perform a first order perturbation theory in $g_1/g_0\sim\Delta/\sqrt{\delta_c^2+\kappa^2}$, Eq.~\eqref{eq:secondorderdiff} becomes a Mathieu equation~\cite{Kovacic:2018} with $V_1(t)\approx\gamma_0$ and $\Delta[\Delta-2V_0(t)]\approx\omega_\mathrm{res}^2-8\Delta\delta_cg_0g_1/[\delta_c^2+\kappa^2]\cos(\omega t)$. Here, we have introduced the time-independent damping
\begin{align}
\gamma_0=\frac{4\kappa\delta_c\Delta g_0^2}{[\delta_c^2+\kappa^2]^2}, \label{gamma0}
\end{align}
and resonance frequency in Eq.~\eqref{omegares}.
The Mathieu equation without damping is known to exhibit instabilities around the parametric resonances {$n\omega=2\omega_{\mathrm{res}}$}~\cite{Kovacic:2018}. In presence of damping $\gamma_0$, instabilities require sufficiently strong driving, provided by the time-periodic term~\cite{Kovacic:2018}. Accordingly we observe in Fig~\ref{Fig:1} superradiance close to the resonance condition $n\omega=2\omega_{\mathrm{res}}$ for pump power in the sidebands $\propto g_1/g_0$ above a certain threshold. This finding is in agreement with previous works where dynamical superradiance has been connected to the Mathieu equation~\cite{Chitra:2015,Molignini:2018}, which again shows that the atomic quantum theory in Eq.~(\ref{eq:effmaster}) gives the correct behavior without describing explicitly the cavity.  Moreover, this allows us to obtain simple results for the  damping rate~\eqref{gamma0} and resonance frequency~\eqref{omegares} and enables us to calculate the threshold in $g_1$. For this we perform a perturbative analysis around the first instability at $\omega=2\omega_{\mathrm{res}}$ reported in the SM~\cite{SM}. We show that the instability occurs at
\begin{align}
g_1^{c}=\frac{2\kappa\omega_{\mathrm{res}}g_0}{\delta_c^2+\kappa^2}.\label{g1c}
\end{align}
The result given by Eq.~\eqref{g1c} is visible as gray solid line in Fig.~\ref{Fig:1}. It agrees well with the threshold found using Floquet theory at $\omega=2\omega_\mathrm{res}$ and its dependence on $\kappa$ highlights the effect of dissipation.

\begin{figure}
	\center
	\includegraphics[width=1\linewidth]{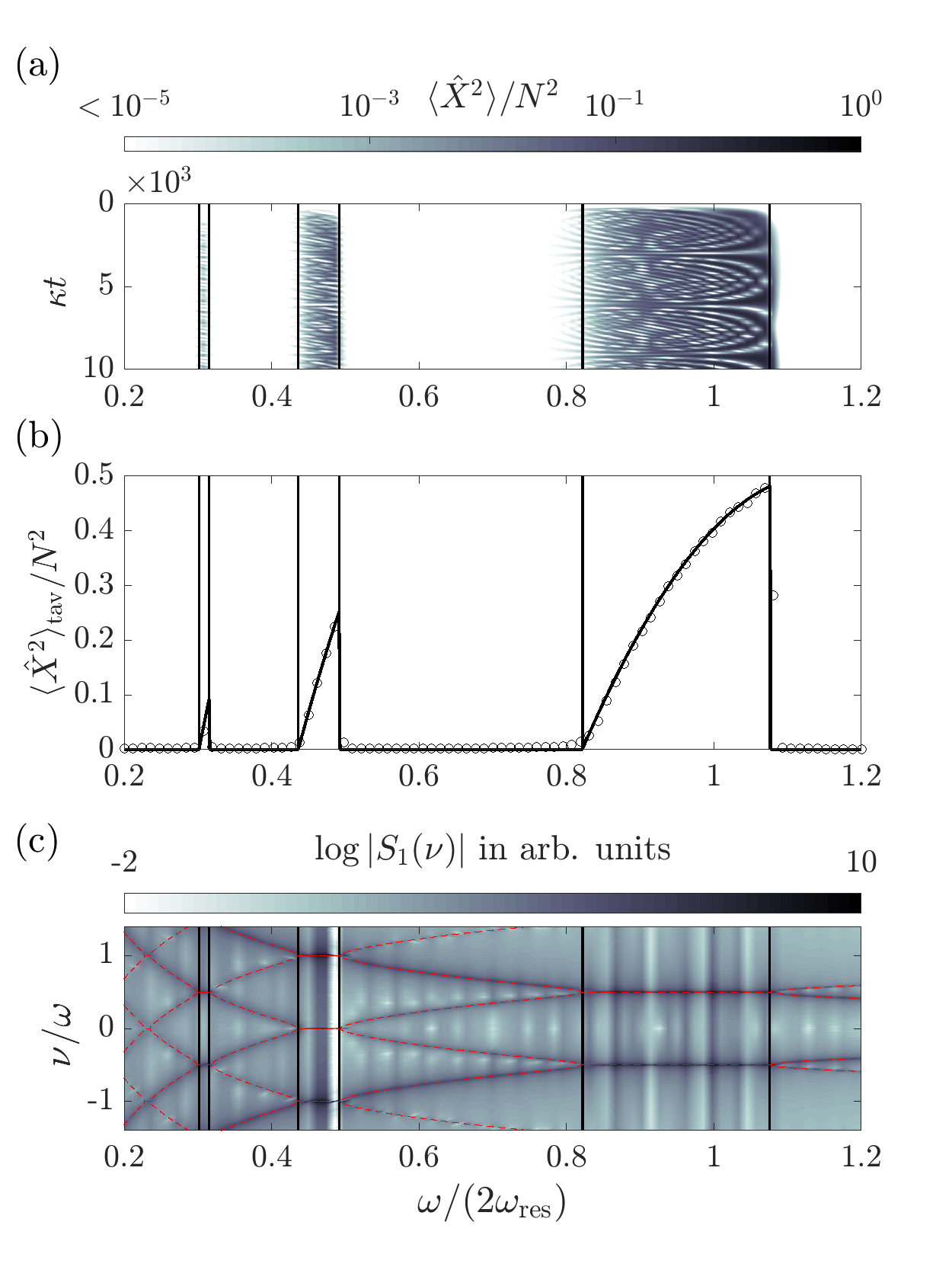}
	\caption{(a) Superradiance order parameter $\langle \hat{X}^2\rangle$ obtained from stochastic simulations as function of time in units of $1/\kappa$ and of $\omega/(2\omega_{\mathrm{res}})$. The vertical black solid lines mark the threshold of superradiance obtained from the atom-only stability analysis. (b) Time-averaged superradiance order parameter $\langle \hat{X}^2\rangle_{\mathrm{tav}}$ as function of $\omega/(2\omega_{\mathrm{res}})$ evaluated after a time $\kappa t=10^4$. Solid black lines (black circles) are obtained from the mean-field (semiclassical) simulations.  (c) Spectrum $S_1(\nu)$ calculated from stochastic simulations as function of $\nu/\omega$ and $\omega/(2\omega_{\mathrm{res}})$ with $\kappa t_0=5\times10^3$ and $t_{\mathrm{max}}=t_0$. The red dashed lines in (c)  are $\nu_{\mathrm{Fl}}$. Remaining parameters are $\delta_c=\kappa$, $\Delta=0.1\kappa$, $g_0=0.5g_c$. The simulations are averaged over $10^4$ trajectories.  \label{Fig:3}}
\end{figure}

\textit{Dynamical response of the atoms}-- For a more comprehensive test of the atom-only theory and the resulting  Floquet theory, we turn to semiclassical simulations of
atoms and cavity (see SM~\cite{SM}) for the same parameters as in Fig.~\ref{Fig:1}, $g_1=0.75g_0$, and different values of $\omega$ (black dashed line in Fig.~\ref{Fig:1}). For these parameters we expect three superradiant regimes around the parametric resonance $\omega=2\omega_{\mathrm{res}}$, $2\omega=2\omega_{\mathrm{res}}$, and $3\omega=2\omega_\mathrm{res}$. We use the semiclassical simulations to calculate $\langle\hat{X}^2(t)\rangle$ shown as function of time $t$ and driving frequency $\omega$ in Fig.~\ref{Fig:3}(a). The atom only theory predicts the borders of the superradiant regimes (black vertical lines), which fully agrees with large values $\langle\hat{X}^2\rangle\propto N^2$. 
In Fig.~\ref{Fig:3}(b), to compare the stationary state, we show the time-averaged superradiant order parameter $\langle\hat{X}^2\rangle_{\mathrm{tav}}$ for both, the semiclassical simulation of atoms and field (circles) and the atom-only mean-field theory (black solid line) at steady state. Both show a remarkable asymmetry of $\langle\hat{X}^2\rangle_{\mathrm{tav}}$ where at each resonance the lower threshold to superradiance is a continuous transition while the upper threshold is marked by a sudden jump of $\langle\hat{X}^2\rangle_{\mathrm{tav}}$. Our semiclassical simulations provide a powerful tool in its own right for studying the dynamics.  The perfect quantitative agreement with the atom-only model demonstrates that the non-trivial microscopic derivation of an effective quantum damping mechanism can be used far from equilibrium. This agreement paves the way for studies of atomic quantum correlations and gaps as well as more involved time-dependent protocols with the full quantum model.

To understand the dynamical response of the atoms, which is crucial to determine whether one finds a subharmonic and time-crystalline order, we employ the two-time correlation function $C_1(t,t_0)$ and calculate its Fourier transform $S_1(\nu)=\int_0^{t_\mathrm{max}}dte^{i\nu t}C_1(t,t_0)$. Here, $t_0$ is a long time after which the dynamics of the system becomes independent of its initial condition and $t_\mathrm{max}$ is a long-time cut-off. The numerical result of $S_1(\nu)$ is shown in Fig.~\ref{Fig:3}(c) as function of $\nu$ and driving frequency $\omega$. The spectrum $S_1(\nu)$ spikes in $\nu$ for all values of $\omega$ that highlight resonances in the atomic medium. These resonances are in  agreement with the Floquet frequencies $\nu_\mathrm{Fl}$ that are visible as red dashed lines in Fig.~\ref{Fig:3}(c). We find $\nu_\mathrm{Fl}=n\omega/2$ in the dynamical superradiant phase corresponding to the parametric resonance $n\omega=2\omega_\mathrm{res}$. This implies that the response of the atoms is flat with respect to the driving frequency which highlights its robustness. Moreover, the response is subharmonic whenever $n$ is odd which becomes clear when considering that the underlying model is a single-mode theory of $\varphi_{\uparrow}$ that oscillates with $\omega_{\mathrm{res}}=n\omega/2$.

\textit{Conclusions}--In conclusion, we have derived and verified an atom-only theory for the time-periodic dissipative Dicke model. With this theory we studied the onset of superradiance including the dynamical response and the threshold determined by the cavity-generated dissipation, the driving frequency and amplitude. Besides the numerical efficiency and maybe most remarkably, this atom-only theory allows us also to describe the long-time relaxation into the DTC that we can understand from an effective cooling mechanism. We remark that all studied quantities in this paper including the superradiance order parameter and spectrum can be measured from the cavity output.  Future theoretical avenues that build on the presented theory could use the atom-only theory to derive quantum fluctuations and low energy excitations of the DTC. This can be used to determine if the emergent states are quantum entangled~\cite{Mattes:2023}. In addition, one can apply the atom-only theory to more complicated systems with many and eventually infinitely many cavity modes. This paves the way to the efficient theoretical description of the atomic medium under periodic driving, which can be used to analyze the generation of squeezed and entangled atomic states with quantum information and metrology applications~\cite{Reilly:2023:2,Luo:2024}.
\begin{acknowledgements}
SBJ acknowledges stimulating discussions with A. Pelster, R. Betzholz, G. Morigi, J. Reilly, and M. J. Holland.
We acknowledge support from Research Centers of the Deutsche Forschungsgemeinschaft (DFG): Projects A4 and A5 in SFB/Transregio 185: “OSCAR”. 
\end{acknowledgements}

	\newpage 

\setcounter{equation}{0}
\setcounter{figure}{0}
\renewcommand*{\citenumfont}[1]{S#1}
\renewcommand{\thesection}{S~\arabic{section}}
\renewcommand{\thesubsection}{\thesection.\arabic{subsection}}
\makeatletter 
\def\tagform@#1{\maketag@@@{(S\ignorespaces#1\unskip\@@italiccorr)}}
\makeatother
\makeatletter
\makeatletter \renewcommand{\fnum@figure}
{\figurename~S\thefigure}
\onecolumngrid
\newpage
\begin{center}
	\textbf{\large Supplemental Material: Dissipative Dicke time crystals: an atoms' point of view}
\end{center}
\vspace{1cm}
\twocolumngrid
\tableofcontents
	\section{Elimination of the cavity field}
In this section we present details on the derivation of the operator $\hat{\beta}$ which is determined by
\begin{align}
	i\frac{\partial\hat{\beta}}{\partial t}=(\delta_c-i\kappa)\hat\beta+\frac{g(t)}{\sqrt{N}}\left(\sx \right)+[\Delta \hat{n}_\uparrow,\hat{\beta}].
\end{align}
To solve this equation we make the ansatz
$\hat\beta(t) = c_+(t) \hat{b}_{\uparrow}^{\dag}\hat{b}_{\downarrow}^{\phantom{\dag}} + c_-(t) \hat{b}_{\downarrow}^{\dag}\hat{b}_{\uparrow}^{\phantom{\dag}}$ which results in the following equations
\begin{align}
	i\frac{dc_{\pm}}{dt}=(\delta_c\pm\Delta-i\kappa)c_\pm+\frac{g(t)}{\sqrt{N}}.
\end{align}
and can be formally solved 
\begin{align}
	c_\pm\approx\frac{-i}{\sqrt{N}}\int_0^t \,d\tau\,e^{\left[-i(\delta_c\pm\Delta)-\kappa\right]\tau}g(t-\tau)
\end{align}
where we dropped the homogeneous solution since it is negligible after times $t\gg\kappa^{-1}$. Assuming $\omega\ll \kappa,\delta_c$ we can use that $g(t)$ changes sufficiently slow such that we can approximate in the integral $g(t-\tau)=g(t)-\tau \dot{g}(t)$  and arrive at
\begin{align}
	c_\pm\approx&-\frac{g(t)}{\sqrt{N}\left[\delta_c\pm\Delta-i\kappa\right]}-\frac{i\dot{g}(t)}{\sqrt{N}\left[(\delta_c\pm\Delta)-i\kappa\right]^2}\nonumber\\
	\approx&-\frac{1}{\sqrt{N}}\left(\frac{g(t)}{\delta_c-i\kappa}+\frac{i\dot{g}(t)}{[\delta_c-i\kappa]^2}\mp
	\frac{\Delta{g}(t)}{[\delta_c-i\kappa]^2}\right)\label{eq:finalcpm}
\end{align}
where we used $\kappa t\gg1$ for the first equation and $\Delta,\omega\ll\delta_c,\kappa$ in the second equation. This shows the result presented in the Letter.
\section{Floquet theory of Lindblad master equations \label{App:FloquetLindblad}}
In this section we provide details on how we calculate the Floquet eigenmodes and frequencies of the Lindblad master equations~(1) and (6) in the main text.

For a time-periodic master equation
\begin{align}
	\frac{\partial\hat{\rho}}{\partial t}=\mathcal{L}(t)\hat{\rho},
\end{align}
where $\mathcal{L}$ denotes the linear operator defined by the right-hand side of Eqs.~(1) and (6) in main text and $\hat{\rho}$ is a density matrix describing atom and cavity or only atoms, respectively. We first decompose the the Liouvillian into a Fourier series
\begin{align}
	\mathcal{L}(t)=\sum_{n=-\infty}^{\infty}\mathcal{L}_ne^{in\omega t}\label{eq:Floquetlindblad}
\end{align}
We then use the standard form of Floquet eigenmodes $\hat{\rho}_\lambda=e^{\lambda t}\hat{\varrho}_\lambda(t)$ with a time-periodic
\begin{align}
	\hat{\varrho}_\lambda(t)=\sum_{n=-\infty}^{\infty}\hat{\varrho}^{(n)}_\lambda e^{in\omega t}
\end{align}
and a complex frequency $\lambda$ with $\mathrm{Re}(\lambda)\leq0$ and $\mathrm{Im}(\lambda)\in[-\omega/2,\omega/2).$
Using the Fouier decomposition and Floquet ansatz we find
\begin{align}
	\lambda\hat{\varrho}^{(n)}_\lambda=\sum_{m=-\infty}^{\infty}\mathcal{L}_m\hat{\varrho}_{\lambda}^{(n-m)}-in\omega\hat{\varrho}_\lambda^{(n)}
\end{align}
This can be reformulated as eigenvalue problem for the larger matrix
\begin{align}
	\boldsymbol{\mathcal{L}}=\sum_{n,m=-\infty}^{\infty}(\mathcal{L}_{m}-in\omega\delta_{m,0})\otimes\ket{n}\bra{n-m}
\end{align}
with eigenvectors 
\begin{align}
	\boldsymbol{\hat{\varrho}}_\lambda=\sum_{n=-\infty}^{\infty}\hat{\varrho}_\lambda^{(n)}\otimes\ket{n}
\end{align}
such that
\begin{align}
	\lambda\boldsymbol{\hat{\varrho}}_\lambda=\boldsymbol{\mathcal{L}}\boldsymbol{\hat{\varrho}}_\lambda.
\end{align}
Whenever we calculate the Floquet eigenmodes we diagonalize $\boldsymbol{\mathcal{L}}$. We remark that in numerical calculations we introduce a cut-off in the Floquet space indexed by $n$. In the above equations we exchange the infinite sums by finite sums
\begin{align}
	\sum_{n=-\infty}^\infty\rightarrow\sum_{n=-M_{\mathrm{cut}}}^{M_{\mathrm{cut}}}.
\end{align}
In our analysis we numerically check that $M_{\mathrm{cut}}$ has been chosen large enough to justify convergence.
\section{Finite size gap for resonant driving}	
In this section we give additional information on how we calculated the dissipative gaps $\gamma$ that are visible in Fig.~2(a) and (b) of the main text. For this we fix $g_0=0.5g_c$ together with (a)~$g_1=0.05g_0$ and (b)~$g_1=0.2g_0$.

In our analysis we take into account that for a finite system size the resonance frequency $\omega_{\mathrm{res}}'\approx\omega_{\mathrm{res}}$ has finite size corrections. Those can be found by diagonalizing $\mathcal{L}_0$ which is the constant component of Eq.~\eqref{eq:Floquetlindblad} for a given particle number $N$. In a first step we therefore calculate $\omega_{\mathrm{res}}'$ as finite size corrected resonance frequency and set $\omega=2\omega_{\mathrm{res}}'$. This allows us to fulfill for every particle number $N$ the parametric resonance condition. 

After we have determined $\omega$ for every value of $N$ the  analysis is based on finding the eigenvalue $\lambda$ of $\boldsymbol{\mathcal{L}}$ as explained in the previous section~\ref{App:FloquetLindblad}. Here we only consider the eigenvalue with $\mathrm{Im}(\lambda)=\omega/2$ such that $-\mathrm{Re}(\lambda)=\gamma$ is the smallest gap. The obtained values of $\gamma$ are visible in Fig.~2(a) and (b) for different cut-offs $M_{\mathrm{cut}}=3$(2,4) as black circles (light gray crosses, gray stars). In Fig.~2(a), for $g_1=0.05g_0$, we find that all three cut-offs predict a very similar value of $\gamma$. For $g_1=0.2g_0$, visible in Fig.~2(b), we find that the smalles cut-off predicts a bending and deviates from the $M_{\mathrm{cut}}=3$. This shows the importance of increasing the cut-off for larger particle numbers. The data set obtained with the larger cut-off $M_{\mathrm{cut}}=4$ is in very good agreement with the data set using $M_{\mathrm{cut}}=3$ which indicates the convergence of the results. 

\section{Comparison of Atom-Cavity and Atom-only dynamics}	
In this section we show the comparison of the atom-cavity with the atom-only dynamics for fast cavity degrees of freedom in the parameter regime $\Delta=0.1\kappa$, $\delta_c=\kappa$.

\begin{figure}
	\center
	\includegraphics[width=1\linewidth]{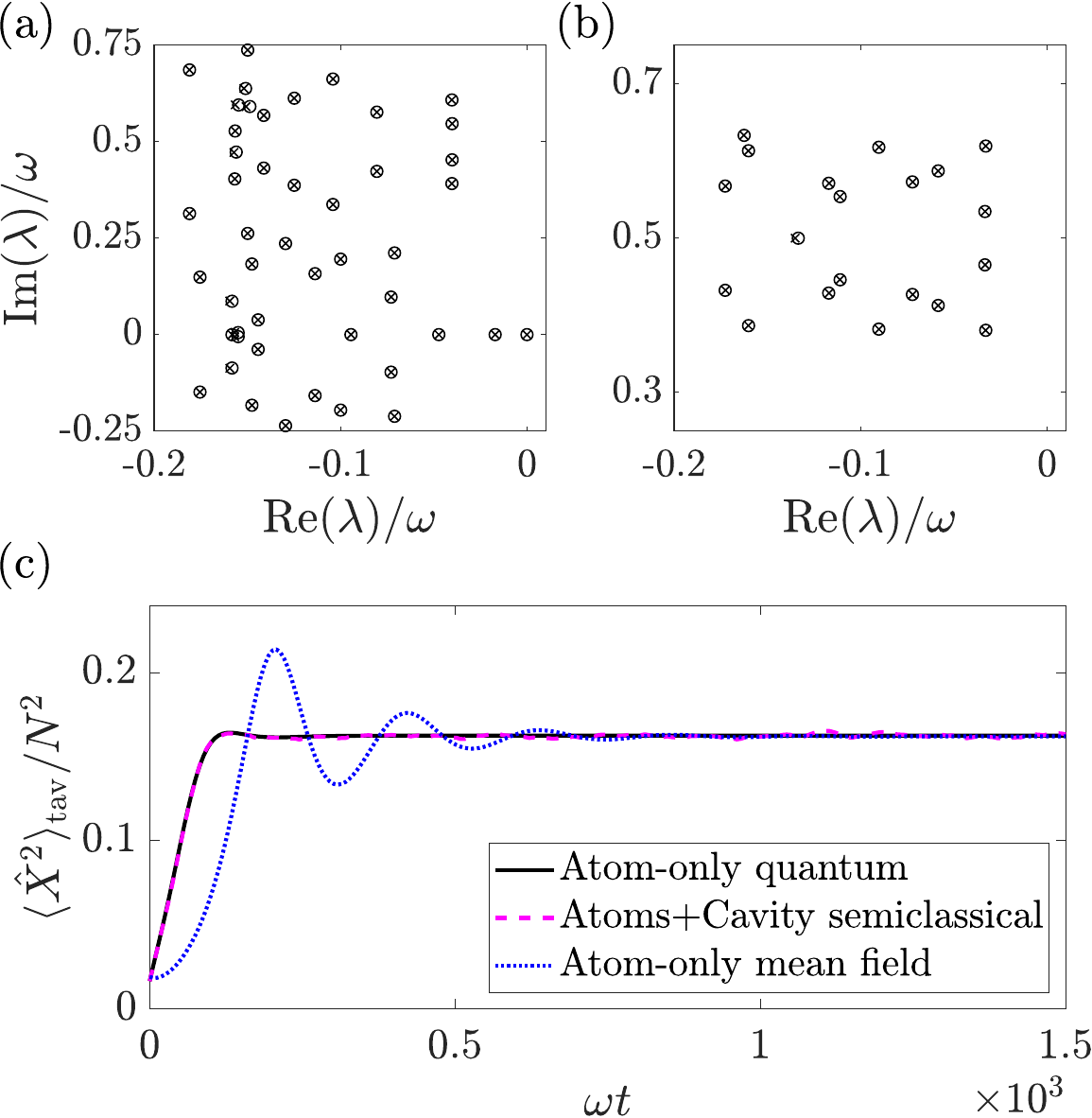}
	\caption{\label{Fig:S1}
		Floquet spectrum of the full atom-cavity Lindbladian visible as circles and the atom-only Lindbladian visible as crosses for (a) $N=10$ and (b) $N=20$. For the calculation of the Floquet spectrum we diagonalized $\boldsymbol{\mathcal{L}}$ with a cut-off $M_{\mathrm{cut}}=4$. For the diagonalization of the full atom-cavity Lindbladian we additionally used a cut-off at the photon number $N_{\mathrm{phot}}=5$.
		(c) The time-averaged observable $\langle\hat{X}^2\rangle_{\mathrm{tav}}$ as function of time for $N=10^2$. The sold line is calculated from the atom-only effective master equation~\eqref{eq:effmaster}, the dashed line from the stochastic semiclassical simulation of atoms and cavity field, and the dotted line from the atom-only mean-field equations~\eqref{eq:phidown},\eqref{eq:phiup}. For all data point we used $g_0=0.5g_c$, $g_1=0.2g_0$, $\omega=2\omega_{\mathrm{res}}$, $\delta_c=\kappa$, and $\Delta=0.1\kappa$. }
\end{figure}

For small atom numbers this comparison is based on the diagonalization of $\boldsymbol{\mathcal{L}}$ using Eq.~(1) in the main text for atoms and cavity and Eq.~(6) for only atoms. 
In  Fig.~S\ref{Fig:S1}(a) we compare the full spectra for $N=10$, $g_0=0.5g_c$, $g_1=0.2g_0$ and $\omega=2\omega_\mathrm{res}$. The circles are obtained by diagonalizing the full Lindbladian describing atoms and cavity and the crosses are calculated using the atom-only Lindbladian. We find very good agreement of these low frequency modes. This indicates that the atom-only description is valid for long timescales. For larger atom numbers it becomes particular hard to diagonalize the full atom-cavity Lindbladian, however, using sparse matrices we were able to compare the twenty lowest frequency modes close to $\mathrm{Im}(\lambda)=\omega/2$ for $N=20$. This comparison is visible in Fig.~S\ref{Fig:S1}(b) where we find very good agreement. For all results we used a cut-off in Floquet space of $M_{\mathrm{cut}}=4$ and a photon cut-off of $N_{\mathrm{phot}}=5$. We checked that our results are converged for these cut-offs. 

For even larger atom numbers we need to rely on semiclassical simulations if we want to treat atoms and cavity on equal footing. Details on how this simulation technique works can be found later in this Supplemental Materials. In order to compare the simclassical dynamics with the full quantum atom-only dynamics we calculate the time-averaged order parameter
\begin{align}
	\langle\hat{X}^2\rangle_{\mathrm{tav}}=\frac{1}{T}\int_{t}^{t+T}d\tau\,\langle\hat{X}^2(\tau)\rangle
\end{align}
as function of time after initializing all atoms in the ground state $\ket{\downarrow}$. The results of the semiclassical analysis is visible as purple dashed line in Fig.~S\ref{Fig:S1}(c) and the atom-only quantum description as solid black line. We choose a parameter set which is in the time-crystalline phase $g_0=0.5g_c$, $g_1=0.2g_0$ and $\omega=2\omega_{\mathrm{res}}$. The atom numbers is $N=100$. We find exceptional agreement of the two simulation methods highlighting that the atom-only theory captures the correct dynamics. The excellent agreement on short and on long timescales is a clear indication that the atom-only theory is valid and we have included the correct cavity-mediated interactions and dissipation. Note that it is far from trivial that the atomic state can evolve towards a dissipative time crystal without explicitly simulating the cavity degrees of freedom.  

\begin{figure}
	\center
	\includegraphics[width=1\linewidth]{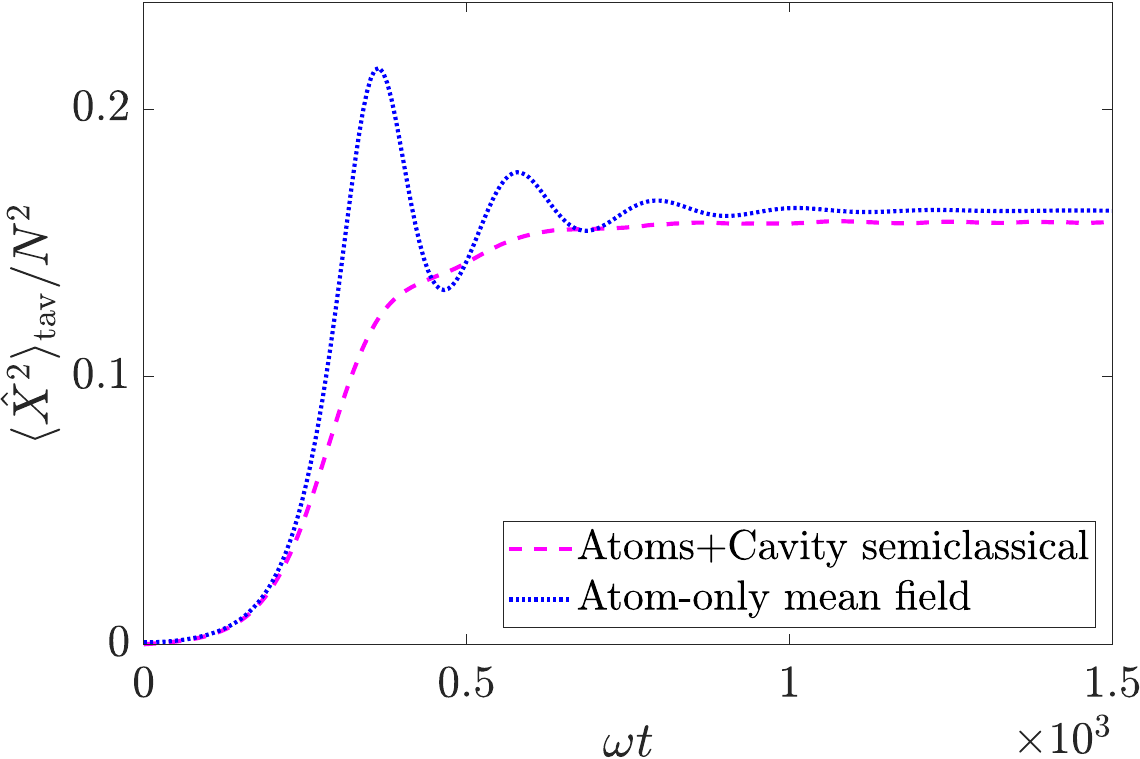}
	\caption{\label{Fig:S2}
		The time-averaged observable $\langle\hat{X}^2\rangle_{\mathrm{tav}}$ as function of time for $N=10^4$. The dashed line is obtained from stochastic semiclassical simulations of atoms and cavity field, and the dotted line from the atom-only mean-field equations~\eqref{eq:phidown},\eqref{eq:phiup}. For all data point we used $g_0=0.5g_c$, $g_1=0.2g_0$, $\omega=2\omega_{\mathrm{res}}$, $\delta_c=\kappa$, and $\Delta=0.1\kappa$. }
\end{figure} 
We can also compare the atom-only mean-field dynamics described by Eqs.~(8) and (9) in the main text. The results of the numerical integration with the same parameters is visible	dotted blue line in Fig.~S\ref{Fig:S1}(c). We observe that the mean-field theory (dotted) evolves to the correct steady state for large times even for a modest value of $N=100$. 	This is an important finding since the simple differential equations~(8) and (9) can be used to observe the relaxation towards a non-trivial stable dynamical state, which maps the whole phase diagram very efficiently in terms of the stationary mean field value $\langle\hat{X}^2\rangle_{\mathrm{tav}}/N^2$. On short timescales we find an exponential growth and oscillations of the mean-field $\langle\hat{X}^2\rangle_{\mathrm{tav}}/N^2$ in Fig.~S\ref{Fig:S1}(c) that is not visible in the quantum simulation of Eq.~\eqref{eq:effmaster}. The oscillations in the mean field value of $\langle\hat{X}^2\rangle_{\mathrm{tav}}/N^2$ depend on the initial value and disappear in the ensemble average.  In Fig.~S\ref{Fig:S2} we also compare the semiclassical method with the mean-field method for very large atom numbers $N=10^4$. Here, we see that the semiclassical simulations predicts the same short-time behavior as the mean-field simulations for large $N$. 
\section{Derivation of the mean-field equations}
In this section we derive the mean-field equations that are presented in the main text from the master equation
\begin{align}
	\frac{\partial \hat{\rho}_{\mathrm{at}}}{\partial t}=-i\left[\hat{H}_{\mathrm{at}}\,\hat{\rho}_\mathrm{at}\right]-\kappa\left(\hat{\beta}^\dag\hat{\beta}\hat{\rho}_{\mathrm{at}}+\hat{\rho}_{\mathrm{at}}\hat{\beta}^\dag\hat{\beta}-2\hat{\beta}\hat{\rho}_\mathrm{at}\hat{\beta}^\dag\right).\label{eq:effmaster}
\end{align}
To do this we derive the differential equations for $\varphi_s=\langle\hat{b}_s\rangle=\mathrm{Tr}(\hat{b}_s\hat{\rho}_{\mathrm{at}})$, $s=\uparrow,\downarrow$. Employing the cyclic property of the trace we obtain 
\begin{align}
	\frac{d\langle\hat{b}_s\rangle}{dt}=-i\left\langle\left[\hat{b}_s,\hat{H}_\mathrm{at}\right]\right\rangle-\kappa\left\langle\hat{\beta}^\dag[\hat{\beta},\hat{b}_s]+[\hat{b}_s,\hat{\beta}^\dag]\hat{\beta}\right\rangle.
\end{align}

Now using the form of $\hat\beta(t) = c_+(t) \hat{b}_{\uparrow}^{\dag}\hat{b}_{\downarrow}^{\phantom{\dag}} + c_-(t) \hat{b}_{\downarrow}^{\dag}\hat{b}_{\uparrow}^{\phantom{\dag}}$ and factorizing higher moments in the bosonic operators we find 
\begin{align}
	\frac{d\varphi_\downarrow}{dt}
	=&-i\frac{g(t)}{\sqrt{N}}\mathrm{Re}\left(c_++c_-\right)|\varphi_\uparrow|^2\varphi_\downarrow\nonumber\\
	&-i\frac{g(t)}{\sqrt{N}}\left(c_+^*+c_-\right)(\varphi_\uparrow)^2\varphi_\downarrow^*\nonumber\\
	&+\kappa(|c_-|^2-|c_+|^2)|\varphi_\uparrow|^2\varphi_\downarrow,\label{eq:phidown}\\
	\frac{d\varphi_\uparrow}{dt}
	=&-i\Delta\varphi_\uparrow-i\frac{g(t)}{\sqrt{N}}\mathrm{Re}\left(c_++c_-\right)|\varphi_\downarrow|^2\varphi_\uparrow\nonumber\\
	&-i\frac{g(t)}{\sqrt{N}}\left(c_++c_-^*\right)(\varphi_\downarrow)^2\varphi_\uparrow^*\nonumber\\
	&+\kappa(|c_+|^2-|c_-|^2)|\varphi_\downarrow|^2\varphi_\uparrow.\label{eq:phiup}
\end{align}

Using Eq.~\eqref{eq:finalcpm} we can then derive
\begin{align}
	\frac{g(t)}{\sqrt{N}}\mathrm{Re}(c_++c_-)=&-\frac{V_0}{N}\\
	\frac{g(t)}{\sqrt{N}}(c_+^*+c_-)=&-\frac{V_0+iV_1}{N}\\
	\kappa(|c_-|^2-|c_+|^2)=&\frac{V_1}{N}
\end{align}
with
\begin{align}
	V_0=&\frac{2\delta_cg^2(t)}{\delta_c^2+\kappa^2}-\frac{4\delta_c\kappa g(t)\dot{g}(t)}{[\delta_c^2+\kappa^2]^2},\\
	V_1=&\frac{4\delta_c\Delta\kappa  g^2(t)}{[\delta_c^2+\kappa^2]^2}.
\end{align}
Inserting the above equations in Eqs.~\eqref{eq:phidown} and \eqref{eq:phiup} leads to the mean-field description shown in the main text. 
\section{Formation of a limit cycle in presence of dissipation}
In this section we demonstrate that Eqs.~(8) and (9) in the manuscript can describe the dynamics into a limit cycle. We also show that in absence of dissipation, $V_1=0$, this set of equations does not describe the stabilization of the limit cycle. This highlights the role of dissipation.

\begin{figure}[h!]
	\center
	\includegraphics[width=1\linewidth]{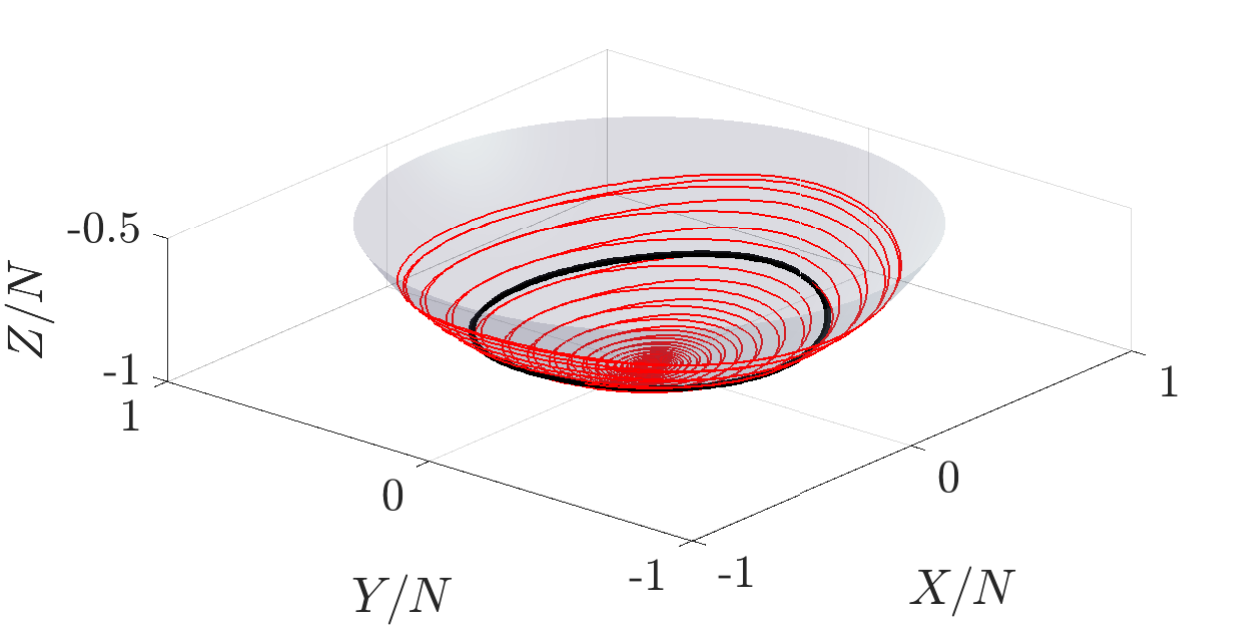}
	\caption{\label{Fig:S3} Mean-field trajectory of $X(t),Y(t),Z(t)$ for $1.1\times 10^3\leq\omega t\leq 1.7\times10^3$calculated from Eqs.~\eqref{eq:phidown},\eqref{eq:phiup} with $V_1\neq0$ (black) and $V_1=0$ (red) and $g_1/g_0=0.2$. For all plots in this figure we used $\delta_c=\kappa$, $\omega=2\omega_\mathrm{res}$, $g_0/g_c=0.5$, and $\Delta=0.1\kappa$.}
\end{figure}  

In Fig.~S\ref{Fig:S3} we show the trajectory of $X(t)=\varphi_\uparrow^*\varphi_\downarrow+\varphi_\downarrow^*\varphi_\uparrow$, $Y(t)=i(\varphi_\downarrow^*\varphi_\uparrow-\varphi_\uparrow^*\varphi_\downarrow)$, $Z(t)=|\varphi_\uparrow|^2-|\varphi_\downarrow|^2$ with $V_1\neq0$ (black) and $V_1=0$ (red) and after times where we have reached the stationary state. Note that $X^2+Y^2+Z^2=N^2$ is a conserved quantity such that trajectories lie always on a sphere. The black curve shows a limit cycle. In contrast, the red curve, where we have switched off dissipation, shows a highly oscillatory behavior. This key finding demonstrates the role of dissipation in providing a cooling mechanism for the atoms which stabilizes the dissipative time crystal.  Consequently, Eqs.~(8) and (9) define a minimal set of equations that describe the dynamics into time-crystalline structures and provides us an efficient method to map out the whole phase diagram shown in Fig.~1 in the main text.

\section{Dynamics in the $|\varphi_\uparrow|\ll \sqrt{N}$ approximation}
In this section we present a special case of the mean-field equations using the approximations $|\varphi_\uparrow|\ll \sqrt{N}$ and $|\varphi_\downarrow|\approx \sqrt{N}$. From these equations we also show how one can derive the second order differential equation that is given in the Letter.

Using $\varphi_\downarrow\approx \sqrt{N}$ we can derive a complex differential equation for $\varphi_\uparrow$ which reads
\begin{align}
	\frac{d\varphi_\uparrow}{dt}=-i\left[\Delta-(V_0+iV_1)\right]\varphi_\uparrow+i(V_0-iV_1)\varphi_\uparrow^*.
\end{align}
This differential equation together with its complex conjugate can be given as a non-hermitian Schrödinger equation
\begin{align}
	\frac{d\boldsymbol{\varphi}_\uparrow}{dt}=-i\boldsymbol{\Upsilon}{\bf H}_{\mathrm{nh}}\boldsymbol{\varphi}_\uparrow. \label{varphidyn}
\end{align}
with $\boldsymbol{\varphi}_\uparrow=(\varphi_\uparrow,\varphi_\uparrow^*)^T$,  the diagonal matrix $\boldsymbol{\Upsilon}=\mathrm{diag}(1,-1)$ and the non-hermitian Hamiltonian
\begin{align}
	{\bf H}_{\mathrm {nh}}=&\begin{pmatrix}
		\Delta-V_0-iV_1&-V_0+iV_1\\
		-V_0-iV_1&\Delta-V_0+iV_1
	\end{pmatrix}.
\end{align}

In a next step we derive from Eq.~\eqref{varphidyn} the second order differential equation that is presented in the main text. For this we define $x_\uparrow=\varphi_\uparrow+\varphi_\uparrow^*$ and $y_\uparrow=i(\varphi_\uparrow-\varphi_\uparrow^*)$ and derive
\begin{align}
	\frac{dx_\uparrow}{dt}=&-\Delta y_\uparrow,\label{eq:x}\\
	\frac{dy_\uparrow}{dt}=&(\Delta-2V_0)x_\uparrow-2V_1y_\uparrow.\label{eq:y}
\end{align}
Taking the derivative of Eq.~\eqref{eq:x} and inserting Eq.~\eqref{eq:y} we arrive at the second order differential equation that was presented in the main text.
\section{Details of the mean-field Floquet analysis}
\label{App:Floquet}
In this section we present details on the Floquet analysis presented in the paper. We use Floquet theory for the linear differential equation~\eqref{varphidyn} with $g=g_0+g_1\cos(\omega t)$ and decompose its square into Fourier components, $g^2(t)=g_0^2+2g_0g_1\cos(\omega t)+g_1^2\cos^2(\omega t)$.
Accordingly, we decompose the linear operator ${\bf A}(t)=-i\boldsymbol{\Upsilon}{\bf H}_{\mathrm{nh}}(t)$ into Fourier components
\begin{align}
	{\bf A}(t)=\sum_{m=-2}^{2}{\bf A}^{(m)}e^{im\omega t}
\end{align}
with
\begin{align}
	{\bf A}^{(0)}=
	\begin{pmatrix}
		-i(\Delta-V_0^{(0)})-V_1^{(0)}&iV_0^{(0)}+V_1^{(0)}\\
		-iV_0^{(0)}+V_1^{(0)}&i(\Delta-V_0^{(0)})-V_1^{(0)}
	\end{pmatrix}
\end{align}
and
\begin{align}
	V_0^{(0)}=&\frac{2\delta_c\left[g_0^2+\frac{g_1^2}{2}\right]}{\delta_c^2+\kappa^2},\\
	V_1^{(0)}=&\frac{4\kappa\delta_c\Delta\left[g_0^2+\frac{g_1^2}{2}\right]}{[\delta_c^2+\kappa^2]^2}.
\end{align}
The rotating components with $\omega$ are given by
\begin{align}
	{\bf A}^{(\pm 1)}=\begin{pmatrix}
		iV_0^{(\pm 1)}-V_1^{(\pm 1)}&iV_0^{(\pm 1)}+V_1^{(\pm 1)}\\
		-iV_0^{(\pm 1)}+V_1^{(\pm 1)}&-iV_0^{(\pm 1)}-V_1^{(\pm 1)},
	\end{pmatrix}
\end{align}
and
\begin{align}
	V_0^{(\pm 1)}=&\frac{2\delta_cg_0g_1}{\delta_c^2+\kappa^2}\mp\frac{i2\delta_c\kappa\omega g_0g_1}{[\delta_c^2+\kappa^2]^2},\\
	V_1^{(\pm 1)}=&\frac{4\kappa\delta_c\Delta g_0g_1}{[\delta_c^2+\kappa^2]^2}.
\end{align}

The next order, rotating with $2\omega$, are given by
\begin{align}
	{\bf A}^{(\pm 2)}=\begin{pmatrix}
		iV_0^{(\pm 2)}-V_1^{(\pm 2)}&iV_0^{(\pm 2)}+V_1^{(\pm 2)}\\
		-iV_0^{(\pm 2)}+V_1^{(\pm 2)}&-iV_0^{(\pm 2)}-V_1^{(\pm 2)}
	\end{pmatrix},
\end{align}
where we defined
\begin{align}
	V_0^{(\pm 2)}=&\frac{2\delta_c\frac{g_1^2}{4}}{\delta_c^2+\kappa^2}\mp\frac{i4\delta_c\kappa\omega \frac{g_1^2}{4}}{[\delta_c^2+\kappa^2]^2},\\
	V_1^{(\pm 2)}=&\frac{4\kappa\delta_c\Delta \frac{g_1^2}{4}}{[\delta_c^2+\kappa^2]^2}.
\end{align}

We apply now Floquet theory to the linear system
\begin{align}
	\frac{d{\bf v}}{dt}={\bf A}{\bf v}.
\end{align}
For this we write $\boldsymbol{\varphi}_\uparrow(t)=e^{\lambda_{\mathrm{Fl}} t}{\bf u}(t)$ with a time-periodic ${\bf u}(t)={\bf u}(t+T)$ and $T=2\pi/\omega$. We find then
\begin{align}
	\lambda_{\mathrm{Fl}}{\bf u}+\frac{d{\bf u}}{dt}={\bf A}{\bf u}.
\end{align}
Since ${\bf u}$ is periodic with period $T$, we can decompose it into Fourier components
\begin{align}
	{\bf u}=\sum_{n}{\bf u}_ne^{in\omega t}
\end{align}
and find
\begin{align}
	\lambda_{\mathrm{Fl}}{\bf u}_n=\sum_{m=-2}^{2}[{\bf A}^{(m)}-in\delta_{m,0}\omega]{\bf u}_{n-m}.
\end{align}
Defining now the $2\times 2$ identity matrix $\mathrm{\bf I}_2$ we can find the solution of the above equation by finding the eigenvalues $\lambda_{\mathrm{Fl}}$ and
eigenvectors
\begin{align}
	{\boldsymbol u}=\sum_{n=-\infty}^{\infty}{\bf u}_n\otimes |n\rangle
\end{align}
of
\begin{align}
	\boldsymbol{A}=\sum_{n=-\infty}^{\infty}\sum_{m=-2}^{2}\left[{\bf A}^{(m)}-in\omega\delta_{m,0}\mathrm{\bf I}_2\right]\otimes|n\rangle\langle n-m|.
\end{align}
The latter is approximated using a sufficiently high cut-off in $n$ such that the results have converged.

\section{Threshold for $\omega=2\omega_{\mathrm{res}}$}
\label{App:Pert}
In this section we derive the estimate for the critical value $g_1^c$ close to resonance $\omega=2\omega_\mathrm{res}+\epsilon$, $\epsilon\ll\omega_\mathrm{res}$, using perturbation theory. This is done in the limit where $g_1\ll g_0$ where we can use the Mathieu equation
\begin{align}
	\frac{d^2x_\uparrow}{dt^2}+2\gamma_0\frac{dx_\uparrow}{dt}+\left[\omega_\mathrm{res}^2-4\Delta b\cos(\omega t)\right]x_\uparrow=0\label{Mathieu}
\end{align}
with $b=2\delta_cg_0g_1/(\delta_c^2+\kappa^2)$. Equivalent derivations of the threshold are given in Ref.~\cite{Taylor:1969,Kovacic:2018}. We assume that $\gamma_0,\sqrt{\Delta b}\ll\omega_\mathrm{res}$ and make the ansatz
\begin{align}
	x_\uparrow=\sum_{n=-\infty}^\infty x_n(t)e^{-i\frac{n\omega}{2}t}.
\end{align}	
The functions $x_n$ are time-dependent Fourier amplitudes. We can then derive a differential equation for $x_n$ that takes the form
\begin{align}
	&\ddot{x}_n-\left(in\omega-2\gamma_0\right)\dot{x}_n-\left[\left(\frac{n^2\omega^2}{4}-\omega^2_\mathrm{res}\right)+in\gamma_0\omega\right] x_n\nonumber\\
	&=2\Delta b(x_{n-2}+x_{n+2})
\end{align}
From the equation above we observe that all amplitudes $x_n$ with $n^2\neq1$ are of higher order in $2\Delta b/\omega^2_\mathrm{res}$. Thus we restrict the equations of motion to $n=\pm1$. The components $x_{\pm1}$ are evolving slowly with $\dot{x}_{\pm 1}/x_{\pm 1}\ll\omega_\mathrm{res}$. In this regime we can approximate the differential equation by
\begin{align}
	\frac{d}{dt}\begin{pmatrix}
		x_1\\
		x_{-1}
	\end{pmatrix}={\bf A}\begin{pmatrix}
		x_1\\
		x_{-1}
	\end{pmatrix}
\end{align}
with
\begin{align}
	{\bf A}=\begin{pmatrix}
		-\gamma_0+i\frac{\epsilon}{2}&i\frac{\Delta b}{\omega_\mathrm{res}}\\
		-i\frac{\Delta b}{\omega_\mathrm{res}}&-\gamma_0-i\frac{\epsilon}{2}
	\end{pmatrix},
\end{align}
where we used $\omega^2\approx4\omega_\mathrm{res}(\omega_\mathrm{res}+\epsilon)$.
The eigenfrequencies are given by the eigenvalues of ${\bf A}$ and to derive them we calculate the characteristic polynomial
\begin{align}
	p(\gamma_{\mathrm{Fl}})=\det\left[{\bf A}-\gamma_{\mathrm{Fl}}{\bf I}_2\right]=(\gamma_{\mathrm{Fl}}+\gamma_0)^2+\frac{\epsilon^2}{4}-\frac{\Delta^2 b^2}{\omega_\mathrm{res}^2}.
\end{align}
The zeros of this polynomial are the eigenfrequencies that are given by
\begin{align}
	\gamma_{\mathrm{Fl}}^{(\pm)}=-\gamma_0\pm\sqrt{\frac{\Delta^2 b^2}{\omega_\mathrm{res}^2}-\frac{\epsilon^2}{4}}.
\end{align}
The fluctuations are stable if the solution is damped. This is the case if
\begin{align}
	g_1^2\leq\frac{4\kappa^2\omega_\mathrm{res}^2g_0^2}{(\delta_c^2+\kappa^2)^2}+\frac{(\delta_c^2+\kappa^2)^2\omega_\mathrm{res}^4}{4\delta_c^2\Delta^2g_0^2}\left(1-\frac{\omega}{2\omega_{\mathrm{res}}}\right)^2.
\end{align}
Therefore, at resonance $\omega=2\omega_\mathrm{res}$, we find the threshold which is given in the main text.
\section{Stochastic Simulation of the dissipative Dicke model}
\label{App:Stoch}
In this section we give the explicit form of the stochastic differential equation that we integrate to find part of the numerical results given in the main text. We define the operators
\begin{align}
	\hat{X}=&\hat{b}^\dag_\uparrow\hat{b}_\downarrow+\hat{b}^\dag_{\downarrow}\hat{b}_\uparrow\\
	\hat{Y}=&i(\hat{b}^\dag_{\downarrow}\hat{b}_\uparrow-\hat{b}^\dag_\uparrow\hat{b}_\downarrow)\\
	\hat{Z}=&\hat{b}_\uparrow^\dag\hat{b}_\uparrow-\hat{b}_\downarrow^\dag\hat{b}_\downarrow.
\end{align}
Our starting point are the Heisenberg-Langevin equations for these operators and cavity degrees of freedom
\begin{align}
	\frac{d\hat{a}}{dt}=&-\kappa\hat{a}-i\delta_c\hat{a}-i\frac{g}{\sqrt{N}}\hat{X}+\sqrt{2\kappa}\hat{a}_{\text{in}}(t),\\
	\frac{d\hat{X}}{dt}=&-\Delta \hat{Y},\\
	\frac{d\hat{Y}}{dt}=&\Delta \hat{X}-2\frac{g(t)}{\sqrt{N}}(\hat{a}+\hat{a}^{\dag})\hat{Z},\\
	\frac{d\hat{Z}}{dt}=&2\frac{g(t)}{\sqrt{N}}(\hat{a}+\hat{a}^{\dag})\hat{Y}.
\end{align}
We introduced the noise operators $\hat{a}_{\text{in}}$ with vanishing mean value $\langle\hat{a}_{\text{in}}\rangle=0$, second moments $\langle\hat{a}_{\text{in}}(t)\hat{a}_{\text{in}}(t')\rangle=0=\langle\hat{a}^{\dag}_{\text{in}}(t)\hat{a}_{\text{in}}(t')\rangle$ and $\langle\hat{a}_{\text{in}}(t)\hat{a}^{\dag}_{\text{in}}(t')\rangle=\delta(t-t')$.

Instead of evolving the complex field we define the hermitian quadratures $\hat{a}_x=\hat{a}^{\dag}+\hat{a}$ and $\hat{a}_p=i(\hat{a}^{\dag}-\hat{a})$. Their time evolution coupled to the spin degrees is given by
\begin{align}
	\frac{d\hat{a}_x}{dt}=&-\kappa\hat{a}_x+\delta_c\hat{a}_p+\sqrt{2\kappa}\hat{\mathcal{N}}^x(t),\\
	\frac{d\hat{a}_p}{dt}=&-\kappa\hat{a}_p-\delta_c\hat{a}_x-2\frac{g(t)}{\sqrt{N}}\hat{X}+\sqrt{2\kappa}\hat{\mathcal{N}}^p(t),\\
	\frac{d\hat{X}}{dt}=&-\Delta\hat{Y},\\
	\frac{d\hat{Y}}{dt}=&\Delta\hat{X}-2\frac{g(t)}{\sqrt{N}}\hat{a}_x\hat{Z},\\
	\frac{d\hat{Z}}{dt}=&2\frac{g(t)}{\sqrt{N}}\hat{a}_x\hat{Y},
\end{align}
with $\hat{\mathcal{N}}^x=[\hat{a}_{\text{in}}(t)+\hat{a}_{\text{in}}^{\dag}(t)]$ and $\hat{\mathcal{N}}^p=-i[\hat{a}_{\text{in}}(t)-\hat{a}_{\text{in}}^{\dag}(t)]$. The stochastic differential equations that are used in the main part of the paper are now derived by exchanging the quantum operators with  real functions using a symmetric ordering. In addition we exchange the quantum noise by classical noise which warrants the correct second moments~\cite{Domokos:2001}. The stochastic semiclassical differential equations are
\begin{align}
	\frac{da_x}{dt}=&-\kappa a_x+\delta_c a_p+\sqrt{2\kappa}{\mathcal{N}}^x(t),\\
	\frac{da_p}{dt}=&-\kappa a_p-\delta_ca_x-2\frac{g}{\sqrt{N}}X+\sqrt{2\kappa}{\mathcal{N}}^p(t),\\
	\frac{dX}{dt}=&-\Delta Y,\\
	\frac{dY}{dt}=&\Delta X-2\frac{g}{\sqrt{N}}a_xZ,\\
	\frac{dZ}{dt}=&2\frac{g}{\sqrt{N}}a_xY,\label{Eq1}
\end{align}
with  $\mathcal{N}^a$ fulfilling $\langle \mathcal{N}^a\rangle=0$ and $\langle \mathcal{N}^a(t)\mathcal{N}^b(t')\rangle=\delta_{ab}{\delta(t-t')}$. We use these stochastic differential equations to simulate the semiclassical dynamics of the coupled spin and cavity dynamics. In this paper we consider as the initial state with $\langle Z\rangle=-N$, $\langle X\rangle=0=\langle Y\rangle$ and the cavity in the vacuum state $\langle a_x\rangle=\langle a_p\rangle=0$. To incorporate quantum fluctuations in the stochastic semiclassical variables $a_x$, $a_p$, $X$, $Y$, and $Z$, we initialize them by independent Gaussian random variables with $\langle a_x^2\rangle=1=\langle a_p^2\rangle$, $\langle X^2\rangle=N=\langle Y^2\rangle$, and $\langle Z^2\rangle=\langle Z\rangle^2=N^2$. With these initial conditions we integrate the stochastic differential equations and average over several initializations that we report in the captions of the figures in the main text. 
\bibliography{article.bib}

\end{document}